\documentclass[aps,
prb,
reprint,
twocolumn,
superscriptaddress
]{revtex4-2}

\usepackage{amssymb,amsmath,amsfonts}
\usepackage{graphicx}
\usepackage{float}
\usepackage{natbib}
\usepackage{color}
\usepackage{lmodern} 
\usepackage[colorlinks=true,bookmarks=false,citecolor=blue,urlcolor=blue]{hyperref}
\usepackage{tikz}
\usepackage{wrapfig}

\graphicspath{ {./Figures/} }

\newcommand{\eps}{\varepsilon }

\newcommand{\dyad}[1] {\overset\leftrightarrow{\mathbf{#1}}}

\newcommand{\elm}{electromagnetic }

\begin{document}

\newcommand{\moscow}{
Moscow Center for Advanced Studies, Moscow, 123592, Russia
}

\newcommand{\mipt}{
Center for Photonics and 2D Materials, Moscow Institute of Physics and Technology, Dolgoprudny 141700, Russia
}

\newcommand{\baum}{Bauman Moscow State Technical University, 2nd Baumanskaya Str., 5, Moscow, Russia}

\title{Analysis of stability and near-equilibrium dynamics of self-assembled Casimir cavities}

\author{Mikhail Krasnov}
\affiliation{\mipt}

\author{Arslan Mazitov}
\affiliation{\mipt}

\author{Nikita Orekhov}
\affiliation{\mipt}
\affiliation{\baum}

\author{Denis G. Baranov}
\email[]{baranov.mipt@gmail.com}
\affiliation{\mipt}

\begin{abstract}
Vacuum fluctuations are a fundamental and irremovable property of a quantized electromagnetic field. These fluctuations are the cause of the Casimir effect -- mutual attraction of two electrically neutral metallic plates in vacuum in the absence of any other interactions. For most geometries and materials, Casimir effect is strictly attractive, leading to the only stable equilibrium configuration with merged plates. Recent observation showed, however, that this unavoidable vacuum-induced attraction can be mitigated by the presence of electrostatic repulsion produced by the formation of double electric layers, and a stable equilibrium between two charged metallic plates in a solution of an organic salt can be reached \cite{munkhbat2021tunable}. Here, we study theoretically in details equilibrium configurations and their dynamical behavior in the system of two parallel metallic films coupled by the Casimir and electrostatic interactions. We analyze the effect of various parameters of the system -- such as the salt concentartion and temperature -- on the equilibrium cavity thicknesses, inspect resonant properties of the resulting an-harmonic optomechanical system near equilibrium, and examine its stochastic dynamics under the influence of thermal fluctuations of the environment.
\end{abstract}

\maketitle
\newpage

\section{Introduction}
Vacuum oscillations of the electromagnetic field are a ubiquitous phenomenon that shows up in a variety of physical systems \cite{milonni1988different,riek2015direct}. Eigenmodes of the electromagnetic field in free space can be treated as quantum harmonic oscillators. Each such oscillator is characterized by a vacuum state with a non-vanishing zero-point energy. In a large number of macroscopic scenarios dealing with the optical response of objects and their equilibrium behavior, the characteristics of this vacuum state can be neglected. However, at the nanoscale, the presence of a nontrivial vacuum state of an optical nanostructure can play a decisive role. A telling example is the Casimir effect, which manifests itself as an attraction between two electrically neutral metallic mirrors in free space in the absence of any other interactions \cite{rodriguez2015classical, woods2016materials, rodriguez2011casimir}. Furthermore, the Casimir effect turned out to be ubiquitous for any system featuring a wave behavior.

The conventional Casimir effect was theoretically predicted for two identical ideally conducting metal plates separated by air \cite{casimir1948attraction}, and demonstrated experimentally more than 50 years later \cite{bressi2002measurement}. It was proved that for two electrically neutral mirror-symmetric objects, the corresponding vacuum potential always increases with the distance between the mirrors, leading to an attractive force \cite{kenneth2006opposites}. The attractive property of the Casimir potential in such systems is the physical reason for the aggregation of nanoparticles in colloids \cite{derjaguin1941acta}, and also causes undesirable friction in nano- and micro-mechanical structures.

It was long after the original effect had been discovered when the existence of structures with a repulsive vacuum potential was theoretically predicted. To implement such an interaction, it was proposed to use materials with magnetic or chiral response \cite{kenneth2002repulsive,zhao2009repulsive}. Later, the repulsive Casimir effect was demonstrated experimentally using dielectric media \cite{munday2009measured}. A big step forward has been made in the recent work \cite{zhao2019stable}, wherein a structure with a stable equilibrium mediated by the zero-point fluctuations of the electromagnetic field between two metal plates separated by two media with accurately chosen permittivity dispersions was experimentally demonstrated (see also refs. \cite{esteso2019casimir, esteso2022effect} for theoretical studies of a similar system).
Stable equilibria were shown to exist in systems comprised by objects of finite cross-sections, for example, by dielectric rods placed inside metallic cavities \cite{levin2010casimir,rodriguez2008stable}. 
Recently, it was shown that the simultaneous presence of the Casimir force and the electrostatic interaction between metal mirrors in a liquid can similarly lead to the presence of stable configurations in a planar system \cite{munkhbat2021tunable, schmidt2023tunable}.

In this paper, we study theoretically in details equilibrium configurations and their dynamical behavior in the system of two parallel metallic films coupled by the Casimir and electrostatic interactions, which reproduces the structures studied experimentally in ref. \cite{munkhbat2021tunable}.
To describe the stable configurations of the system, we combine the macroscopic Lifhitz formalism for description of the Casimir force with molecular dynamics simulations allowing for the description of the repulsive electrostatic interaction.
Guided by this combined analytical-numerical model, we analyze the effect of various parameters of the system -- such as the salt concentration and temperature -- on the equilibrium cavity thicknesses.
Next, we  study resonant properties of the resulting mechanical oscillator near equilibrium.
Finally, we simulate the vertical and in-plane stochastic dynamics of the coupled system under the influence of thermal fluctuations of the environment.




\section{Theoretical  model}

The system under study is illustrated in Fig. 1. It incorporates two parallel metallic films of thickness $L$ floating in a water solution of an organic salt.
The characteristic distance between the two films lies in the range of a few hundred nm. There are two main interaction mechanisms that govern the behavior of this system: the electrostatic repulsion due to the formation of electric double layers on the surfaces of the metallic films, and the attractive Casimir effect due to the vacuum oscillations of the \elm field between the metallic films. 
The total potential of the system is given by the sum of these two interactions:
\begin{equation}
    U = U_{C} + U {e}.
    \label{Eq_potential_1}
\end{equation}
In the following we discuss the models and approximations that will allow us to describe these two contributions to the total potential in an analytical or semi-analytical way.

\begin{figure}[t!]
\centering\includegraphics[width=.9\columnwidth]{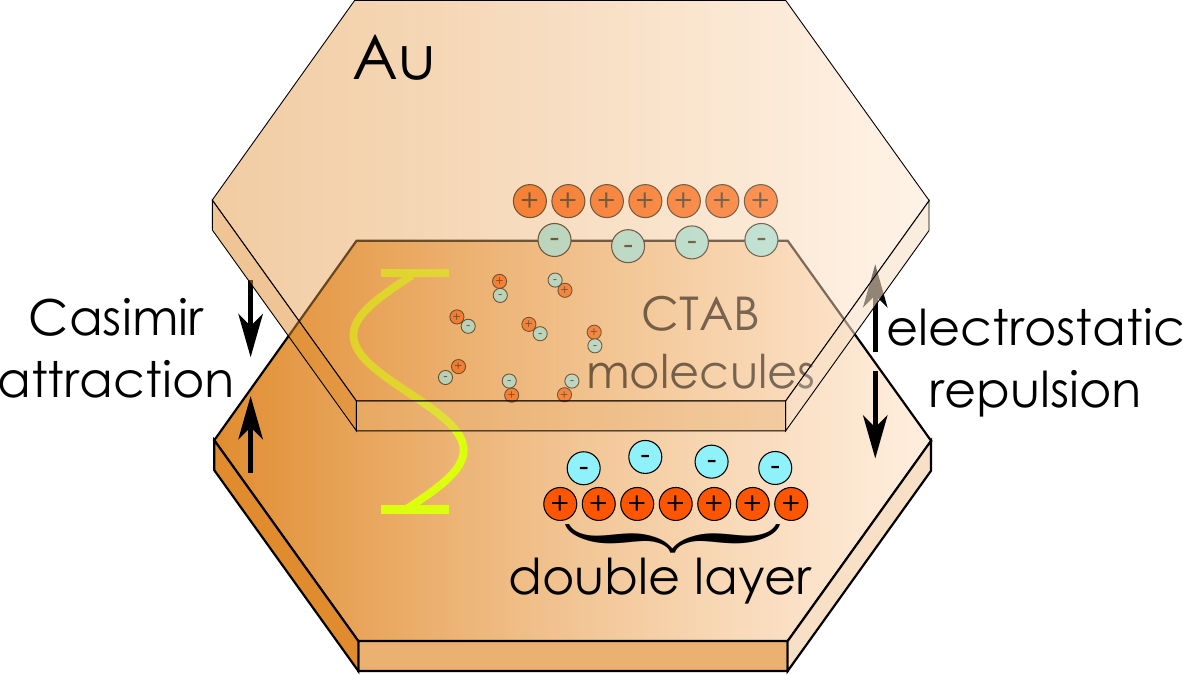}
\caption{Sketch of the system under study. Two parallel metallic films floating in an aqueous solution of an organic salt. 
The films are balance by two interaction: the attractive Casimir force and the repulsive electrostatic force.
The attractive Casimir force is the result of vacuum fluctuations of the electromagnetic field in the region between the metallic films, where optical cavity modes are created. The electrostatic force appears due to the formation of charged electric double layers on the metallic surfaces.}
\label{fig1}
\end{figure}

\subsection{Casimir potential}

The conventional Casimir effect between two parallel surfaces made of a perfect electric conductor (PEC) at a distance $L$ is described by the simple expression \cite{casimir1948attraction}:
\begin{equation}
    U_{C}^{\mathrm{PEC}} = -\frac{\hbar c \pi ^2}{240 L^3},
    \label{Eq_1}
\end{equation}
where $\hbar$ is the reduced Planck constant and $c$ is the speed of light.
In systems incorporating non-perfect conductors and dispersive materials, the Casimir force deviates sufficiently from this textbook expression, and may even become repulsive \cite{kenneth2002repulsive, munday2009measured}. A more general description is offered by the Lifshitz formula \cite{Dzyaloshinskii1961,lifshitz1992theory}. In this framework the Casimir energy is obtained by transforming the oscillating integral over real frequencies to an integral over imaginary frequencies with the use of so called Wick rotation \cite{rodriguez2007virtual,lambrecht2006casimir}. The resulting potential per unit area is given by:
\begin{equation}
    U_{C} = \frac{\hbar}{2\pi}\int_0^{\infty} {d\xi \int {\frac{d^2 \mathbf{k}_{\parallel} }{(2\pi)^2} \ln \det \dyad G } },
    \label{Eq_2}
\end{equation}
where $\mathbf{k}_{\parallel}$ is the real-valued in-plane component of the wave vector in the gap region of thickness $L$, $\dyad G = \mathbb{I} - \dyad{R}_1  \dyad{R}_2 e^{-2 K_0 L}$, and 
\begin{equation}
    \dyad{R}_i = \begin{pmatrix}
    r_i^{ss} & 0\\
    0 & r_{i}^{pp}
    \end{pmatrix}
\label{Eq_3}
\end{equation}
is the reflection matrix for $i$-th side of the system ($i=1,2$); $r_{i}^{q}$ are the Fresnel reflection coefficients for $i$-th subsystem and polarization state $q$ ($q=s,p$) evaluated at imaginary frequency $\omega = i \xi,\ \xi \in \mathbb{R}$. $K_0=\sqrt{\mathbf{k}_{\parallel}^2 + \xi^2/c^2}$ is the $z$-component of the wave vector in the gap between the two mirrors evaluated at imaginary frequency.

The reflection coefficients in Eq. \eqref{Eq_2} must be evaluated at imaginary frequencies. This requires an analytical expression for the permittivity of every component of the planar structure, which (the expressions) admit an analytical continuation into the complex plane. To that end, we describe the permittivity of gold with the Drude model:
\begin{equation}
   \varepsilon_{\text{Au}} = \varepsilon_{\infty} - \frac{\omega_P^2}{\omega(\omega + i \gamma_D)},
   \label{Eq_4}
\end{equation}
where $\eps_{\infty}$ is the background permittivity, $\omega_P$ is plasma frequency, and $\gamma_D$ is the electron collision rate. 
Although in real-frequency calculations the background permittivity $\eps_{\infty}$ is often set to $\eps_{\infty} > 1$ to account for the interband transitions of a metal, we set $\varepsilon_{\infty} = 1$.
This is justified by the fact that the permittivity must asymptotically approach $1$ in the high-frequency limit in order to comply with the basic analytical properties of permittivity and permeability.
Otherwise, the UV contributions to the vacuum energy due to non-transparent mirrors at high frequencies will dominate causing an unphysical result \cite{rodriguez2007virtual,klimchitskaya2009casimir,lambrecht2006casimir}.
Furthermore, we set $\omega_P = 9$ eV, which approximates well the plasma frequency of gold, and set $\gamma_D = 0$ corresponding to a lossless metal, which, surprisingly, yields the best agreement with experimental results \cite{klimchitskaya2009casimir, lambrecht2000casimir}.

\begin{figure*}[hbt!]
\centering\includegraphics[width=1\textwidth]{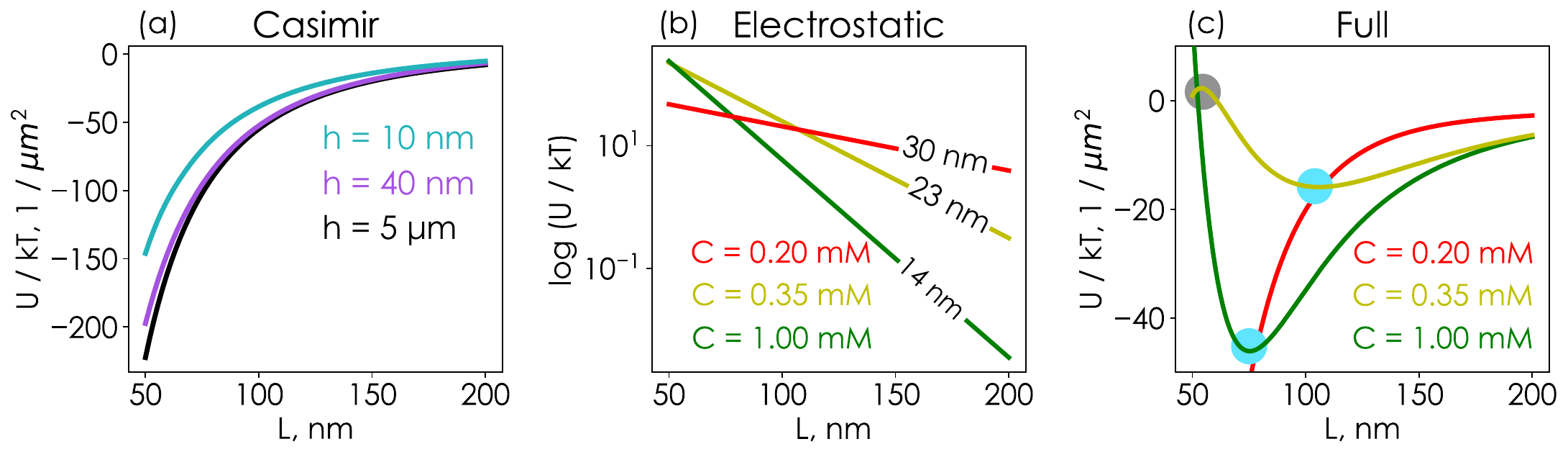}
\caption{\textbf{Casimir and electrostatic potentials in the system under study.} (a) Examples of the Casimir potential per unit area between two gold films of thickness $h$ separated by water evaluated with Lifshitz formula, Eq. \eqref{Eq_1}.
(b) Examples of the electrostatic potential between two gold films evaluated for different values of  CTAB concentration $C$ and assuming the linear surface charge-concentration relationship, Eq. \eqref{Eq_8}. Note the logarithmic scale on the $y$-axis. The value on each curve is the Debye-H\"uckel screening length calculated for the respective concentration. 
(c) Examples of the total potential for 40 nm thick gold films and various CTAB concentrations. While the total potential for smaller CTAB concentration shows a monotonic behavior, for higher concentrations the potential exhibits a series of minima (blue circle) and maxima (grey circles) corresponding to stable and unstable equilibria of the system.
}
\label{fig2}
\end{figure*}

The permittivity of water was described by approximating the experimental data from ref. \cite{segelstein1981complex} with the Debye-Lorentz analytical model:
\begin{equation} \label{Eq:Lor}
    \varepsilon_{\mathrm{H_2 O}}(\omega) = 1
    + \frac{\varepsilon_D - 1 }{1 - i \omega\tau}
    + f \frac{\omega_{0}^2}{\omega_{0}^2-\omega^2 - i\gamma_0 \omega},
\end{equation}
where $\omega_{0} = 18.4$ eV, $f_1 = 0.8$, $\gamma_0 = 13.5$ eV, $\eps_D = 75$ and $ \tau = \text{49000 fs} $.
This approximation reproduces very well water refractive index of $n \approx 1.33$ in the visible, and also correctly describes the high-frequency and low-frequency limits (see Fig. S1 within the Supplemental
Material \cite{SI_ref} for the plot of this approximation).
In our calculations we have assumed that small concentrations of the electrolyte do not change substantially the refractive index of water.

Figure \ref{fig2}(a) presents the resulting Casimir potential between two gold films of thickness $h$ in water as a function of the face-to-face distance $L$ between the films.
All three show monotonous potential growth with the distance.
In the following analysis we are going to stick to $h = 40$ nm thick films corresponding to the typical value of Au films used in ref. \cite{munkhbat2021tunable}.
Fitting the data in the double logarithmic scale (see Fig. S2 within the SM \cite{SI_ref}) reveals that the Casimir potential between Drude films can be approximated by the power law:
\begin{equation}
    U_C = \frac{A}{L^{\beta} }
    \label{Eq_appr_7}
\end{equation}
with $A = -124083$ eV/$\mu$m$^2$ and $\beta = 2.55$. 
We are going to use approximation \eqref{Eq_appr_7} to describes the Casimir potential in the following analysis.

\subsection{Electrostatic potential}

In addition to the Casimir attraction, the metallic films experience electrostatic repulsion. This repulsion is caused by the adsorption of the salt ions (either positive or negative) onto the metallic plates. Oppositely charged ions are electrostatically attracted to the first ionic layer and form the second \textit{diffuse} layer, resulting in screening of the electrostatic potential of the first layer. 
The resulting picture of electrostatic interaction is called an electric double layer (EDL) \cite{helmholtz1853ueber, grahame1947electrical}. 

\begin{figure*}[t!]
\centering\includegraphics[width=1\textwidth]{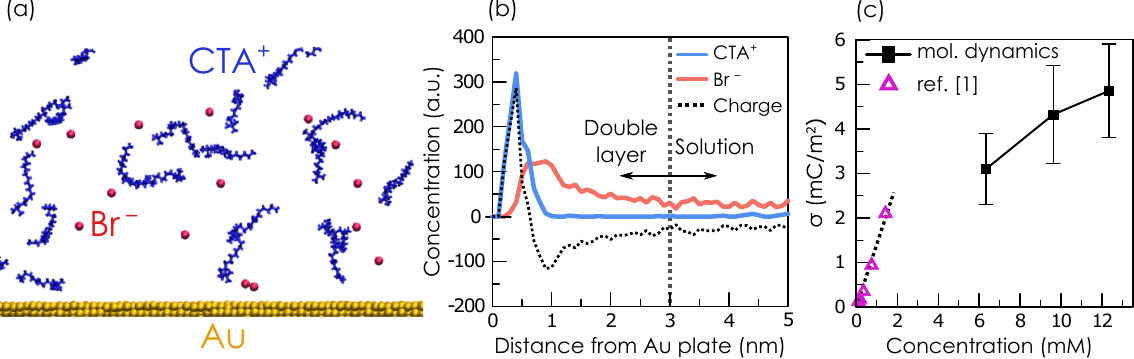}
\caption{\textbf{Molecular dynamics simulation of CTAB adsorption on a gold film.} (a) Snapshot from the molecular dynamics simulation showing the geometry of the computational cell with CTA$^+$ and Br$^-$ ions dissolved in water in the vicinity of an Au film.
(b) Spatial distribution of CTA$^+$ (blue) and Br$^-$ (red) concentrations, as well as total algebraic charge (black dashed) in the computational cell evaluated for CTAB concentration of 9.5 mM.
(c) Surface charge density $\sigma$ as a function of CTAB concentration.
Black markers: the results of our MD simulations. 
Red markers: values obtained from theoretical analysis of experimental data by Munkhbat et al. \cite{munkhbat2021tunable}.
}
\label{MD}
\end{figure*}

One of the most commonly used models to describe EDLs is the Gouy-Chapman model \cite{Schmickler2010, russel_saville_schowalter_1989}. In this model, the ions in the diffuse layer near each plate are assumed to obey Boltzmann statistics as a function of distance $x$ from the plate: 
\begin{equation}
\begin{split}
    n_{+}(x) &= C \exp{(-zq_0\phi(x) / k_B T)},\\
    n_{-}(x) &= C \exp{(zq_0\phi(x) / k_B T)},
\end{split}
\end{equation}
where $n_{+}(x)$ and $n_{-}(x)$ are the densities of the positively and negatively charged ions in the solution, $C$ is the concentration of ions with valence $z$ in a solution, $q_0$ is the elementary charge, $k_B$ is the Boltzmann constant, $T$ is the temperature and $\phi(x)$ is the electrostatic potential. In general, $\phi(x)$ is derived from the the Poisson-Boltzmann equation \cite{Schmickler2010}, which can be expressed in its linearized form using the Debye-H\"uckel approximation $zq_0\phi(x)/k_B T \ll 1$:
\begin{equation}
    \frac{d^2\phi(x)}{dx^2} = \kappa^2 \phi(x),
\end{equation}
where
\begin{equation}
    \kappa = \sqrt{\frac{2 (zq_0)^2 C}{ \varepsilon \varepsilon_0 k_B T}}
    \label{Eq_kappa}
\end{equation}
is the inverse Debye-H\"{u}ckel length, $\varepsilon$ is the static permittivity of the solution, and $\varepsilon_0$ is the vacuum permittivity.  Taking into account the charge balance condition
\begin{equation}
    \int_0^\infty \rho(x) dx = \int_0^\infty zq_0(n_{+}(x) - n_{-}(x)) dx = - \sigma,
\end{equation}
which essentially constrains total surface charge density $\sigma$ to balance the charge density in a solution, a straightforward calculation \cite{Schmickler2010} yields 
\begin{equation}
    \phi(x) = \frac{\sigma}{\varepsilon \varepsilon_0 \kappa} \exp{(-\kappa x)}.
\end{equation}
Finally, the expression for the potential $\phi(x)$ can be used to estimate the electrostatic fields of both plates and to calculate their repulsive force $F_e$ and electrostatic energy $U_e$ per unit area, given the distance $L$ between the plates:
\begin{equation}
\begin{split}
    F_e &= \frac{2 \sigma^2}{\varepsilon \varepsilon_0} \exp{(-\kappa L)}, \\
    U_e &= - \int_L^\infty F_e(L') dL' = \frac{2 \sigma^2}{\varepsilon \varepsilon_0 \kappa} \exp{(-\kappa L)}.
\end{split}
\label{Eq_12}
\end{equation}

The surface charge density $\sigma$ itself is a function of the salt concentration, $\sigma = \sigma(C)$.
In the most general case, this dependence is described by the so called Langmuir isotherm \cite{liu2006some,latour2015langmuir}, which describes adsorption by assuming the adsorbate behaves as an ideal gas. 
The particular implementation in ref. \cite{munkhbat2021tunable} used CTAB (Cetyltrimethylammonium bromide) as soluble organic salt. In the following we are going to analyze the behavior of this particular salt.

To analyze the salt adsorption on the surface in the present system, we performed a series of constant-temperature/constant-pressure molecular dynamics (MD) simulations at $T=300$ K, $P=1$ atm. in a computational cell containing quasi-infinite golden plate, Fig. \ref{MD}(a) (see Appendix A for details). Computational cell with the size of $\approx 160 \times 80 \times 160$ \AA$^3$ contained 65000 water molecules. The number of CTAB complexes varied from 20 to 40. For each CTAB concentration $C$ we performed three independent $500$ ns long MD runs with random initial positions of molecules. 
Figure \ref{MD}(b) shows spatial density distribution of CTA$^+$ and Br$^-$ ions along surface normal for the equilibrated state of the system. 
The total surface charge density $\sigma$ was then evaluated by summing all charges present within the 3 nm region from the surface (dashed line in Fig. \ref{MD}(b)).

Figure \ref{MD}(c) shows the resulting evaluated surface charge density for a series of CTAB concentrations (black markers).
Although we were not able to reach the experimental concentrations of $\approx 1$ mM due to the limitations on the computational cell size, our results reasonably agree with the values obtained in ref. \cite{munkhbat2021tunable} for concentrations below $1.5$ mM (purple markers in Fig. \ref{MD}(c)). 
Munkhbat et al. estimated the surface charge density by fitting the analytical interaction potential of the system (Eq. \eqref{Eq_potential_1}) to match the measured equilibrium distance $L_{eq}$ (extracted from optical reflection measurements), which yields the value of $\alpha$ around $1.5\ \mathrm{mC}/(\mathrm{mM} \cdot \mathrm{m}^2)$. The extracted dependence remains linear up to about 1.5 mM, after which CTAB molecules tend to form so called mycells -- molecular clusters, which qualitatively change the adsorbing behavior of the salt \cite{patel2014ph,lopez2012structural}.

For the sake of this study, we thus adopt the linear model of the surface charge-salt concentration relationship:
\begin{equation}
    \sigma = \alpha C,
    \label{Eq_8}
\end{equation}
with $\alpha = 1\ \mathrm{mC}/(\mathrm{mM} \cdot \mathrm{m}^2)$, which by the order of magnitude agrees with both the previous experimental results, and with our MD simulations.

Figure \ref{fig2}(b) shows the resulting electrostatic potentials calculated according to Eq. \eqref{Eq_12} for different salt concentrations assuming the linear model of the surface charge-concentration relationship, Eq. \eqref{Eq_8}.
The effect of the salt concentration on the electrostatic potential is two-fold. On the one hand, an increase of $C$ proportionally increases the surface charge density $\sigma$. At the same time, it reduces the screening length $1/\kappa$.
As a result, the electrostatic potential acquires larger magnitudes at short distances, but starts to decay faster, Fig. \ref{fig2}(b).

\begin{figure*}[t!]
\centering\includegraphics[width=.8\textwidth]{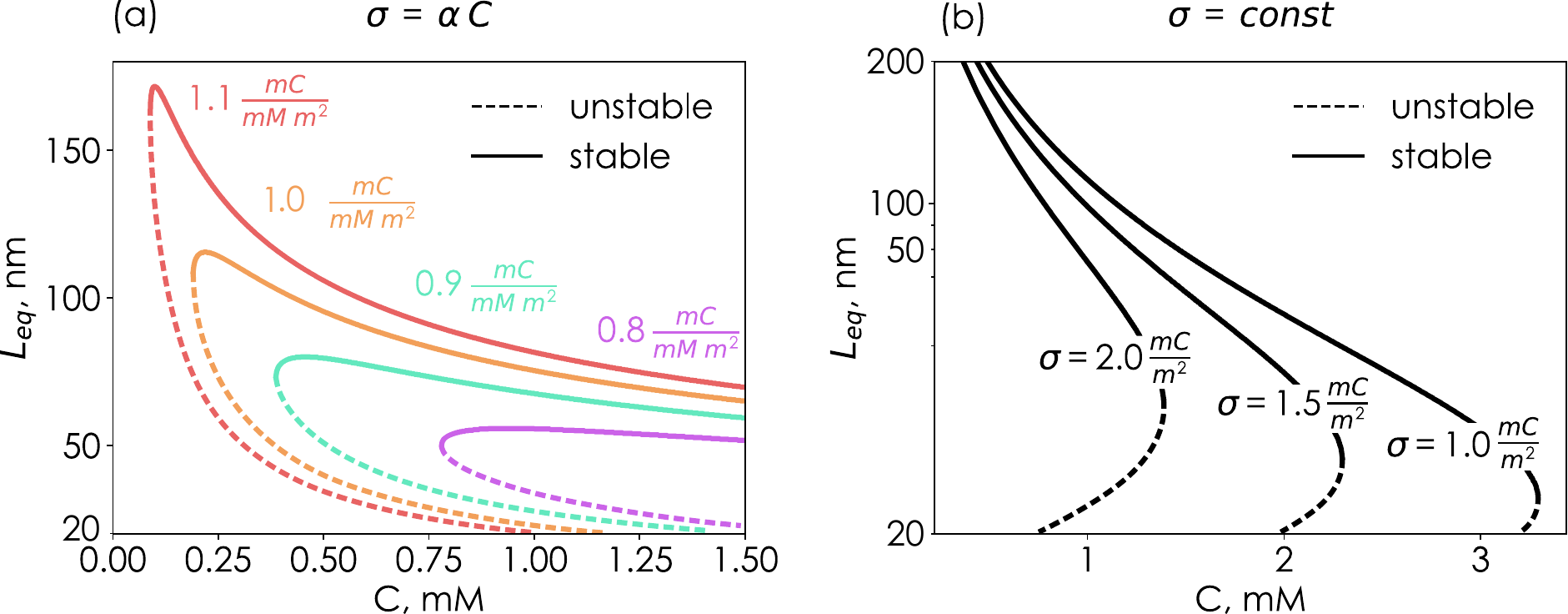}
\caption{\textbf{Evolution of the equilibrium states of the system.} (a) Stable (solid) and unstable (dashed) thicknesses of the self-assembled cavity (corresponding to the local potential minima and maxima, respectively) obtained for the linear salt-charge relationship $\sigma = \alpha C$ for different values of $\alpha$. (b) Same as (a) but obtained for a series of fixed concentration-independent surface charge densities $\sigma = 1,\ 1.5,\ 2.0$ mC/m$^2$. Note the logarithmic scale on the vertical axis of the plot.}
\label{fig3}
\end{figure*}

The total potential is illustrated on Fig. \ref{fig2}(c) for a series of realistic CTAB concentration values.
A peculiar interplay of the two interactions occurs depending on salt concentration. Different functional dependence of the attractive and repulsive potentials makes it possible for a local potential extremum to appear (blue circles). 
For sufficiently small salt concentration the magnitude of the electrostatic repulsion is not sufficient to overcome the Casimir attraction. As a result, the total potential exhibits a monotonic behavior with no extrema ($C = 0.2$ mM, red curve).
For higher salt concentrations, however, the electrostatic repulsion becomes comparable to the Casimir attraction, giving rise to a pair of the potential minimum and maximum, Fig. \ref{fig2}(c), corresponding to a stable and an unstable equilibrium, respectively ($C = 0.35,1$ mM, yellow and green curves).
In the following section, we examine in more details how these equilibria of the total potential evolve with the parameters of the system.

\section{Results}

\subsection{Stable and unstable equilibria of the system}

With these models and approximations for the potential at hand, let us now examine the evolution of the local maxima and minima of the total potential with the salt concentration. 
Figure \ref{fig3}(a) presents the dependence of the stable (solid) and unstable (dashed) equilibrium cavity thickness $L_{eq}$ as a function of the CTAB concentration, $C$, for a series of values of $\alpha$.
The data is limited to the range of CTAB density $C < 1.5 \text{mM}$ because of the aforementioned limited applicability of the linear model $\sigma =  \alpha  C$.
For any given $\alpha$ both the stable and the unstable equilibrium positions vary continuously with $C$ in a certain range.
Whereas the unstable equilibrium exhibits a simple monotonic dependence, the stable equilibrium thickness reaches the highest attainable value at a certain concentration, dependent on $\alpha$.
Furthermore, both equilibria cease to exist below a threshold concentration, which is also determined by $\alpha$. Exactly at that concentration the two equilibria coalesce in a single point, leaving no equilibria for concentrations below the threshold value.

\begin{figure}[b!]
\centering\includegraphics[width=1\columnwidth]{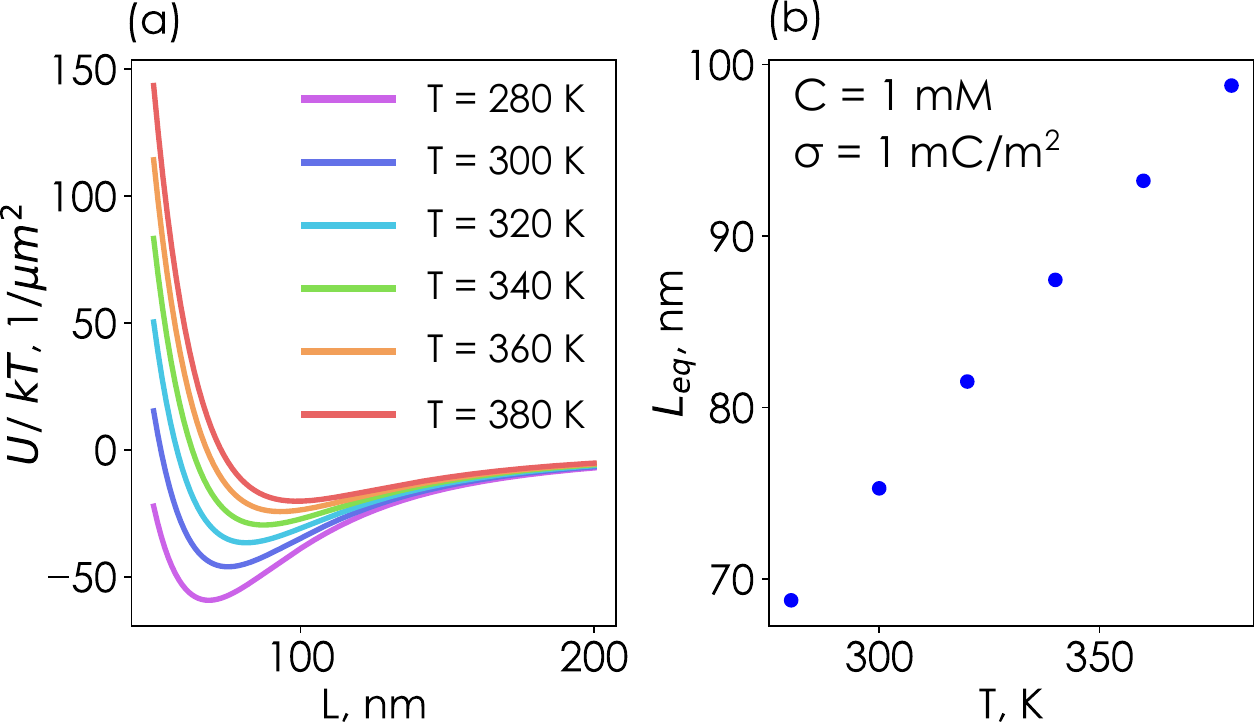}
\caption{\textbf{Temperature effect on the equilibrium configurations of the system.} (a) Total per-unit-area potentials of the self-assembled system  for a fixed CTAB density of $C = 1$ mM for a series of temperature values. (b) The resulting evolution of the  stable thickness of the self-assembled cavity corresponding to the local potential minimum with temperature.}
\label{fig4}
\end{figure}

\begin{figure*}[t!]
\centering\includegraphics[width=.8\textwidth]{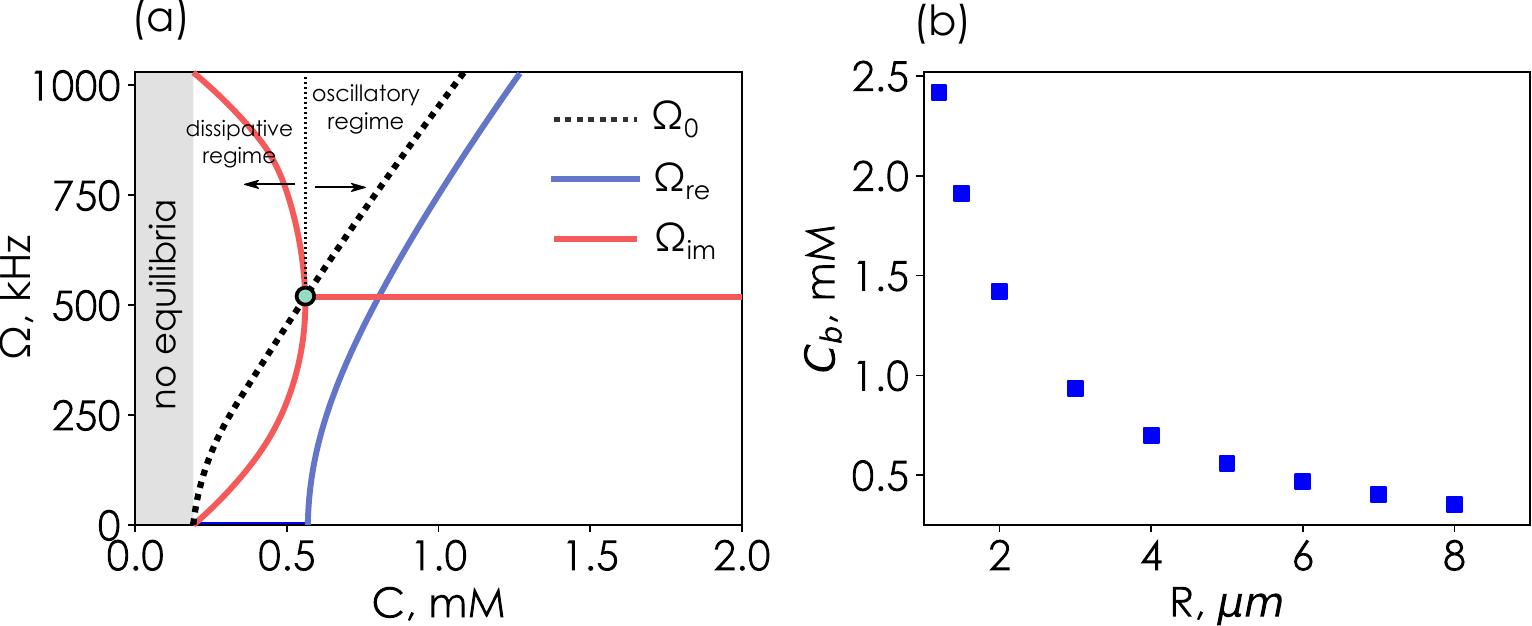}
\caption{\textbf{Mechanical eigenfrequencies of the self-assembled system.} (a) Complex-valued resonant mechanical frequencies of the self-assembled cavity as a function of CTAB concentration for $R=5$ $\mu$m circular flakes. Dashed curve shows the eigenfrequencies in the hypothetical drag-less case, $\gamma_S = 0$. The green circle marks the bifurcation point where the two complex-valued eigenfrequencies coalesce.
(b) Dependence of critical concentration $C_b$ on the flake radius $R$.
}
\label{fig5}
\end{figure*}

For completeness, in addition to the linear model of the surface charge-salt concentration relationship, we briefly examine the existence and behavior of equilibrium configurations assuming a constant surface charge density, $\sigma = \mathrm{const}$. This situation may correspond to the case of saturated surface adsorption, when increasing the concentration of the salt in the solution does not result in further adsorption of ions into the first layer. Alternatively, this may be encountered if the metallic (or non-metallic) films have a fixed internal charge density.

Figure \ref{fig3}(b) shows the resulting dependence of the stable (solid black) and unstable (dashed black) equilibrium distances $L_{eq}$ as a function of the salt concentration, $C$, for a series of surface charge densities $\sigma = 1,\ 1.5,\ 2.0$ mC/m$^2$. Both equilibria exhibit a qualitatively different behavior in comparison to the case of the linear model. 
For each surface charge density both the stable and unstable equilibria show a monotonic behavior and coalesce at a certain concentration, above which no stable or unstable equilibria exist.
In contrast to the case of the linear charge-concentration model, the equilibrium thickness does not reach an upper limit and instead increases indefinitely with decreasing CTAB concentration.

\subsection{Effect of temperature}

Next we examine the effect of temperature on the equilibrium configurations of the system. Variation of temperature has a two-fold effect of the total potential of the system: it affects the Casimir potential, and at the same time affects the Debye-H\"uckel screening length.

At room temperatures $T_{\textrm{room}} \approx 300$K its effect on the Casimir potential is negligible. 
Indeed, the characteristic cavity thickness $L \sim 100$ nm corresponds to the characteristic photon energy of $E \sim \hbar \pi c / (n_{\mathrm{H_2 O}} L) \approx 4$ eV, which is well above the thermal excitation energy at room temperature ($K_B T_{\textrm{room}} \approx 25$ mev). 
Thus, population of the relevant photonic modes of the cavity that mainly contribute to the Casimir potential remains negligible, $e^{- E / (K_B T_{\textrm{room}}) } \ll 1$. 
Thus, the Casimir potential can be calculated as if the system were at zero temperature \cite{Sushkov2011, brevik2006thermal}.

Nevertheless, the electrostatic potential does experience  significant changes even upon moderate temperature changes within the $0 \dots 100 ^\circ$ C range (see Fig. S3 of  Supplemental Material \cite{SI_ref}). Increasing the solution temperature reduces the inverse Debye-H\"{u}ckel screening length $\kappa$, see Eq. \eqref{Eq_kappa}, thus increasing the contribution of electrostatic repulsion to the total potential.

Figure \ref{fig4}(a) shows the combined per-unit-area potential of two 40 nm thick gold films for a series of temperatures in the range from 0 to 100 C for a 1 mM CTAB solution (resulting in $ 1~\mathrm{mC}/\mathrm{m}^2$ surface charge density according to the accepted value of $\alpha = 1~\textrm{mC}/(\textrm{m}^2 \cdot \textrm{mM})$).
This causes the stable equilibrium cavity thickness to shift towards higher values with increasing temperature. At the same time, the depth of the potential minimum reduces with increasing temperature.
Figure \ref{fig4}(b) shows the resulting equilibrium cavity thickness as a function of temperature, which shows almost a linear dependence.

\subsection{Optomechanical properties}

The entire system of two metallic films equilibrated by the joint action of the Casimir and electrostatic interactions represents a mechanical oscillator. Detuning the cavity thickness from the stable equilibrium position in either way causes a restoring force directed back to the equilibrium position.
Next we study the mechanical resonant properties of the system near the stable equilibrium.

For the sake of the following analysis assume the two gold flakes are circular disks of identical size, placed on top of each other, and can only perform motion along the $z$-axis pointing from flake '1' toward flake '2'.
The vertical dynamics of each flake along the $z$ axis in the absence of any external impact can be modelled by the homogeneous Langevin equation \cite{neuman2004optical, volpe2013simulation}:
\begin{subequations}
\begin{align}
    m \Ddot{z}_1  = - \gamma_\perp \dot{z}_1 + F_{1}(z_1, z_2), \label{Eq_Lang_1}\\
    m \Ddot{z}_2  = - \gamma_\perp \dot{z}_2 + F_{2}(z_1, z_2), \label{Eq_Lang_2}
\end{align}
\end{subequations}
where $z_1$ ($z_2$) is the vertical coordinate of the bottom (top) film, $m = \rho S h$ is the mass of the metallic film with $S = \pi R^2$ being the geometric area of a single flake, and $\gamma_\perp$ is the hydrodynamic drag coefficient for vertical motion.
The drag experienced by the flake during its laminar motion in water can be described by Stokes’ law \cite{volpe2013simulation}:
\begin{equation}
    \gamma_\perp = 6 \pi \eta R_S
\end{equation}
with $\eta$ being viscosity of water.
The Stokes’ radius $R_S$ of the object is determined by many factors, and it can be further greatly affected by the close presence of the substrate (the boundary). As a reasonable assumption we estimate the Stokes’ radius of the disk by its geometric radius, $R_S = R$.

Following ref. \cite{munkhbat2021tunable}, we evaluate the total restoring force acting on films '1' and '2' within the overlap approximation by multiplying the per-unit-area potential gradient $dU/dL$ with the geometric area of the film:
\begin{equation}
    F_{1} = - S dU/d L, \quad F_2 = - F_1
\end{equation}
with $L = z_1 - z_2$ being the relative vertical displacement of the films, synonymous with the cavity thickness.
The total potential density per unit area in the vicinity of the equilibrium can be approximated by the harmonic potential:
\begin{equation}
    U = U_0 + \frac{1}{2} k(L - L_{eq})^2 ,
    \label{Eq_quadratic}
\end{equation}
where $L_{eq}$ is the stable cavity thickness for the particular salt concentration and the corresponding surface charge density, and
\begin{equation}
    k = \left. \frac{d^2 U}{d L^2} \right\rvert_{L = L_{eq}}
    \label{Eq_stiff}
\end{equation}
is the stiffness of the quadratic potential. 
For reference, we show in Fig. S4 (see SM \cite{SI_ref}) the exact analytical combined potential of the system, as well as the harmonic (quadratic) approximation near the corresponding equilibrium cavity thickness. 
The resulting potential plots clearly show that the quadratic potential describes the system fairly accurately in within 5-10 nm from the equilibrium, but starts to deviate outside that range.

Subtracting Eq. \eqref{Eq_Lang_2} from \eqref{Eq_Lang_1} we obtain:
\begin{equation}
    m \Ddot{L}  = - \gamma_\perp \dot{L} +  F(L),
\label{dynamic_eq}
\end{equation}
where $F = 2 F_1 \approx -2A k (L - L_{eq})$. The factor $2$ in the latter expression comes from the simultaneous action of the Casimir/electrostatic force on both films.
Looking for a harmonic solution of Eq. \eqref{dynamic_eq} $L (t) = L_{eq} + \delta L e^{i \Omega t}$ we find a pair of eigenfrequencies of the damped mechanical oscillator:
\begin{equation}
    \Omega = \frac{i\gamma_\perp \pm \sqrt{8\pi R^2m k - \gamma_\perp^2}}{2m}.
    \label{Eq_mech_eig}
\end{equation}

Figure \ref{fig5}(a) shows the resulting mechanical eigenfrequencies of the self-assembled cavity as a function of CTAB concentration at room temperature for $R = 5~\mu$m circular flakes.
Below a certain CTAB concentration defined by $\gamma_\perp ^2 - 8\pi R^2 m k > 0$ Eq. \eqref{Eq_mech_eig} yields two purely imaginary frequencies corresponding to dissipate dynamics of the system:
\begin{equation}
    \Omega = \frac{i}{2m} \left( \gamma_\perp \pm 
    \sqrt{\gamma_\perp^2 - 8\pi R^2m k}
    \right).
    \label{Eq_mech_eig_1}
\end{equation}
Above that critical density we obtain a single complex-valued eigenfrequency:
\begin{equation}
    \Omega = \frac{\sqrt{8\pi R^2m k - \gamma_\perp^2 }}{2m}  + 
    i \frac{\gamma_\perp}{2m} .
    \label{Eq_mech_eig_2}
\end{equation}
Dashed line shows the eigenfrequency spectrum corresponding to the hypothetical case of a drag-free system, $\gamma_\perp = 0$:
\begin{equation}
    \Omega_{0} = \sqrt{ \frac{2 k \pi R^2}{m} }.
    \label{Eq_mech_eig_lossless}
\end{equation}


The critical concentration $C_b$, at which the mechanical eigenfrequencies undergo the bifurcation, depends on the film surface area $S$. Figure \ref{fig5}(b) shows the critical concentration as a function of the circular flake radius $R$, revealing a monotonic decreasing dependence. For the typical flake radius of a few microns, the bifurcation concentration $C_b$ lies around the realistic value of 1 mM.

\begin{figure}[t!]
\centering\includegraphics[width =1\columnwidth]{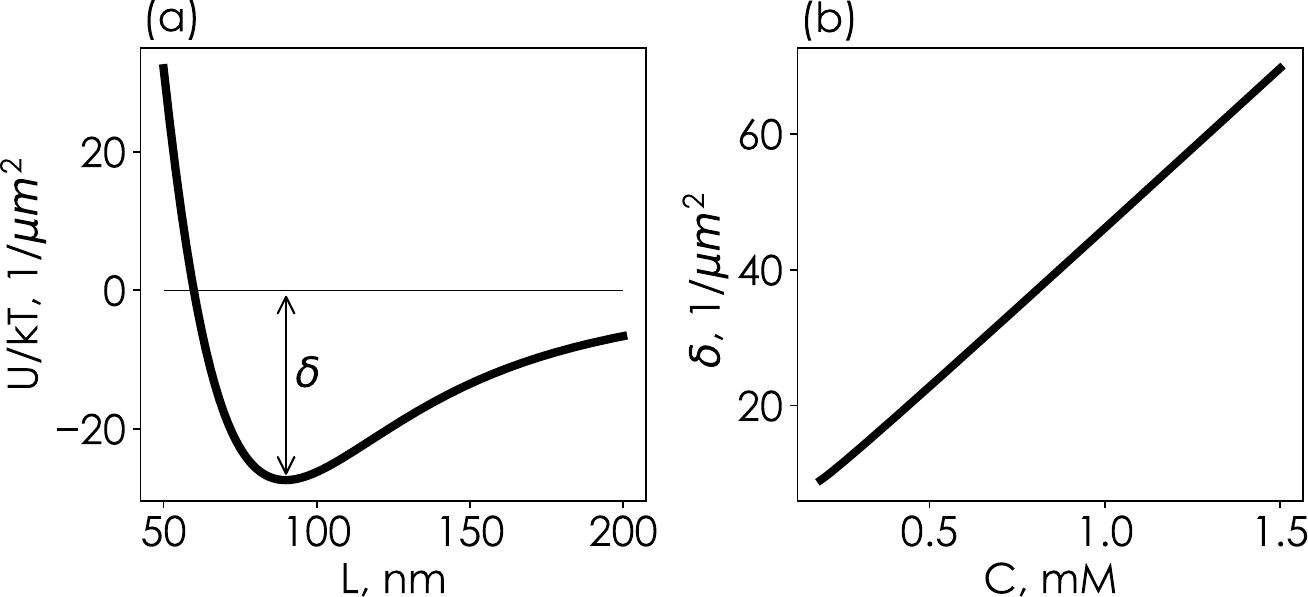}
\caption{\textbf{Anharmonism of the mechanical self-assembled oscillator.} (a) An example of the total per-unit-area potential illustrating the measure of the potential anharmonicity $\delta$. The potential at $L = \infty$ is zero. (b) The dependence of the potential density depth $\delta$ on CTAB concentration at room temperature.}
\label{fig6}
\end{figure}

The analysis performed above clearly indicates that the mechanical oscillator formed by a self-assembled cavity is anharmonic. 
There are many ways in which we could quantify the degree of this anharmonicity.
We choose to quantify the anharmonicity by the depth $\delta$ of the total per-unit-area potential minimum (normalized by $K_B T$ at $T = 300$ K) with respect to the potential value at infinite film-to-film distance, Fig. \ref{fig6}(a).
The smaller the $\delta$ parameter, the more anharmonic the oscillator is. A perfect harmonic oscillator would have an infinite value of $\delta$.
Figure \ref{fig6}(b) shows the potential depth $\delta$ as a function of CTAB concentration (assuming the accepted value of $\alpha$) at room temperature, $T = 300$ K. It reveals a nearly linear dependence of the anharmonicity parameter on CTAB concentration. 


\begin{figure*}[t!]
\centering\includegraphics[width = 1\textwidth]{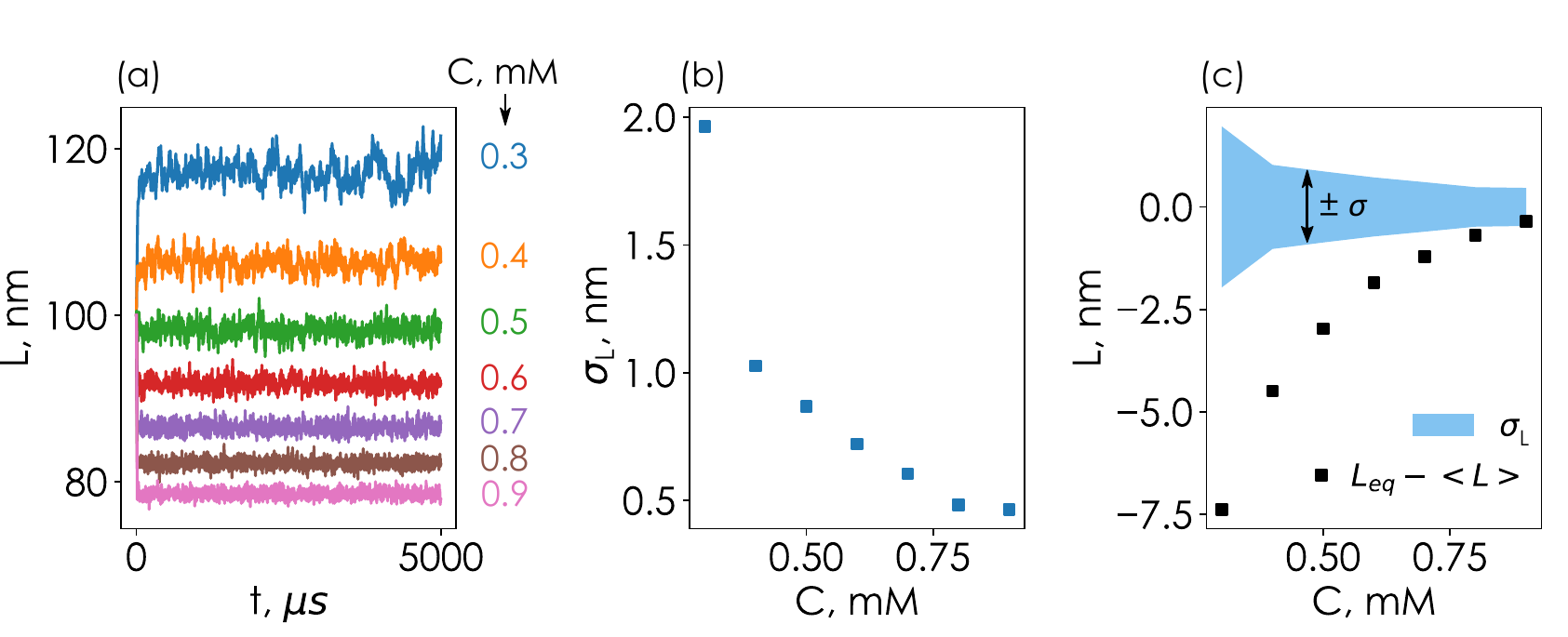}
\caption{\textbf{Dynamics of the self-assembled cavity near the equilibrium under the impact of thermal fluctuations.} (a) Simulated dynamics of the vertical displacement of a self-assembled cavity $L$ for a series of CTAB concentrations (on the right)), all calculated for room temperature, $T=300$ K. The films are assumed to be circular disks of $R = 5~\mu$m radius.
(b) Standard deviation $\sigma_L$ of the cavity thickness $L$ during its stochastic evolution. 
(c) Equilibrium cavity thickness $L_{eq}$ (dots) compared against the the standard deviation range defined by $\langle L \rangle \pm \sigma_L$ (shaded area). The data is centered by subtracting the time-average cavity thickness $\langle L \rangle$ for each CTAB concentration.
}
\label{fig7}
\end{figure*}

\subsection{Stochastic behavior: vertical dynamics}

Next we turn to the analysis of the dynamical behavior of self-assembled cavities in the vertical direction near the equilibrium state under the influence of thermal fluctuations.
The stochastic dynamics of the structure near the equilibrium position can be modelled by the Langevin equation, Eq. \eqref{dynamic_eq} with the added noise term \cite{neuman2004optical, volpe2013simulation}:
\begin{equation}
    m \Ddot{L}  = - \gamma_\perp \dot{L} + F(L) +  
    \sqrt{2 \gamma K_B T} ( f_1(t) + f_2(t)),
\end{equation}
where $f_{1,2}(t)$ are two independent white noise terms with correlation function
\begin{equation}
    \langle f_i(t) f_j(t - \tau) \rangle = \delta(\tau) \delta_{ij},
\end{equation}
and $K_B$ is the Boltzmann constant.
Note that for the purpose of these simulations the exact potential per unit area $U(z)$, and not its quadratic approximation, was used.

\begin{figure*}[t!]
\centering \includegraphics[width=\textwidth]{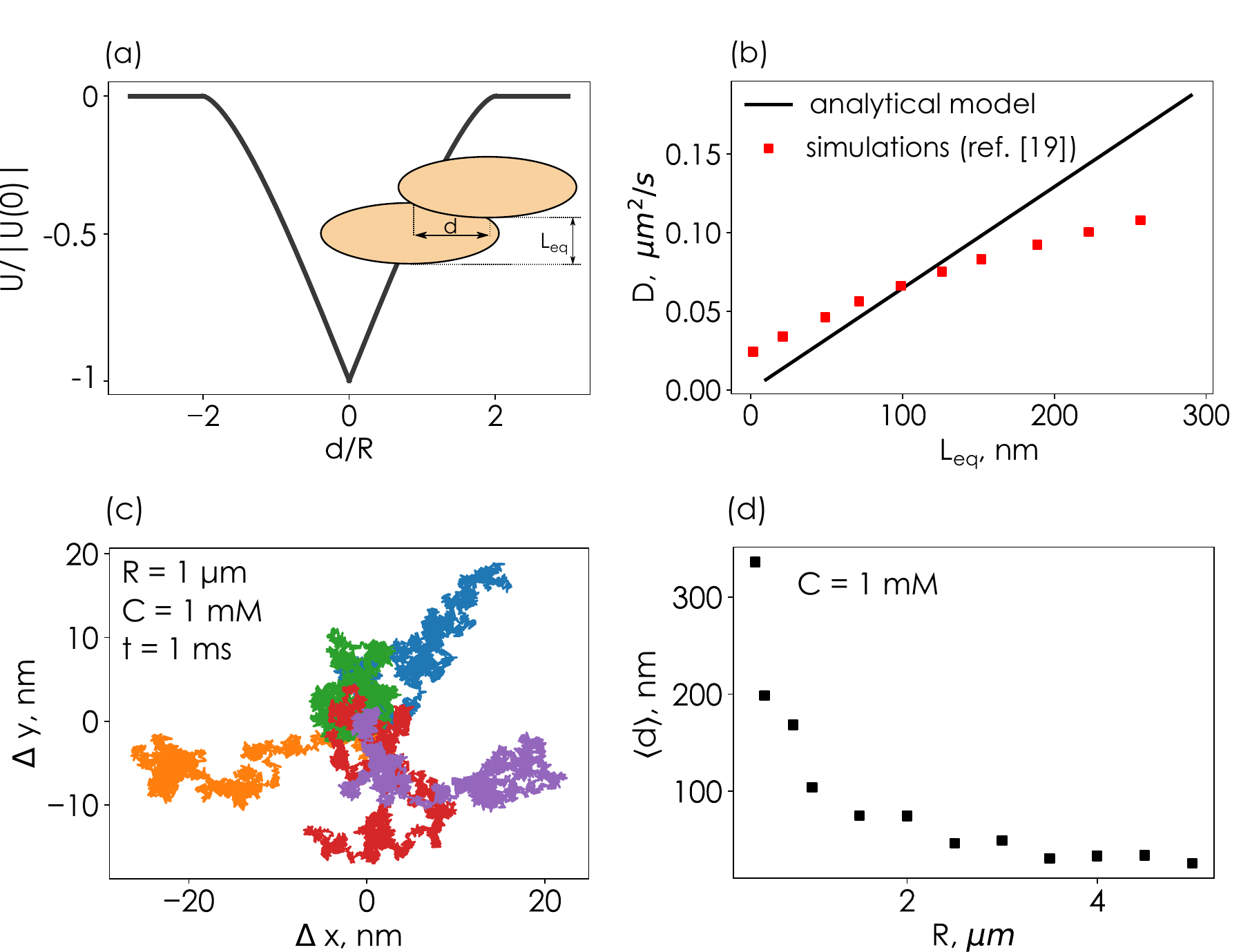}
\caption{\textbf{Translational dynamics of the system in the horizontal plane.}
(a) Potential of two coupled flakes upon in-plane motion  vs center-to-center distance $d$ normalized by the potential at $d=0$. Inset: sketch of the geometry of the system. Two metallic parallel flakes separated by a vertical distance $L_{eq}$ experience translational in-plane motion.
(b) Diffusion coefficient $D$ as a function of the flake-to-flake distance $L$. Line: analytical model, Eq. \eqref{Eq_Diff}. Dots: data from ref. \cite{schmidt2023tunable}.
(c) Simulated in-plane dynamics of a self-assembled cavity for a pair of coupled $R = 1~\mu$m circular flakes in a $C = 1$~mM CTAB solution at room temperature, $T=300$ K. All trajectories begin at $x=y=0$ and span the range of $t=1$~ms.
(d) Time-average center-to-center displacement $\langle d \rangle$ as a function of the disk radius $R$ calculated based on the simulated in-plane trajectories for $C = 1$~mM.
}
\label{fig9}
\end{figure*}

Figure \ref{fig7}(a) presents a number of simulated trajectories of the vertical displacement $L$ of a self-assembled cavity formed by two $R = 5~\mu$m circular flakes for a series of CTAB concentrations performed at $T=300$ K.
After an initial transient regime 
every trajectory reaches its equilibrium position determined by the CTAB concentration, around which it starts to experience Brownian dynamics.
Figure S5 of Supplemental Material shows examples of simulated vertical trajectories for lower CTAB concentrations.


The amplitude of this Brownian dynamics depends strongly on salt concentration. The potential well gets stiffer with increasing $C$, which leads to smaller deviation of the system from the equilibrium position.
This is supported by Fig. \ref{fig7}(b), which shows the standard deviation $\sigma_L$ of the cavity thickness $L$ calculated based on the simulated dynamical trajectories. For each CTAB concentration the standard deviation was evaluated using the data from the part of trajectory after the initial transient regime ($t > 100~\mu$s).
Clearly, the standard deviation gradually reduces with increasing salt concentration.
The resulting values in the range from 0.5 nm to 2 nm are remarkably close to the standard deviation of the self-assembled cavities measured in the original experimental work \cite{munkhbat2021tunable}, which supports the validity of our simple dynamical model.

By comparing the calculated expected value of the cavity thickness $\langle L \rangle$ with the corresponding equilibrium thickness $L_{eq}$ for each salt concentration we arrive at a remarkable conclusion:  
the equilibrium thicknesses $L_{eq}$ (corresponding to the local potential minimum) generally lie outside of the standard deviation range $\langle L \rangle \pm \sigma_L$, Fig. \ref{fig7}(c). 
The data shown in this plot is centered by subtracting the time-averaged cavity thickness $\langle L \rangle$ for each salt concentration.
This behavior can be understood in light of the asymmetry of the total potential around the equilibrium point. As a result, the system spends more time on the less stiff part of the potential ($L > L_{eq}$) than on the other side, which leads in the observed behavior.

\subsection{Horizontal dynamics}

Next we examine translational (in-plane) dynamics of circular flakes in the horizontal plane with respect to each other.
For the sake of this analysis we consider two identical circular flakes of radius $R$. Assume the vertical distance between the flakes is fixed to $L_{eq}$ determined by the salt concentration, $L_{eq} = L_{eq}(C)$.
Let $x_{1,2}$ and $y_{1,2}$ denote the in-plane coordinates of the centers of flakes "1" and "2".
Writing the dynamical equations of the in-plane motion for each flake and subtracting one from another we obtain:
\begin{equation}
\begin{split}
    m \Ddot{x}  = - \gamma_\parallel \dot{x} + F_{x} + \sqrt{2 \gamma K_b T} ( f_{x1} (t) + f_{x2} (t)),\\
    m \Ddot{y}  = - \gamma_\parallel \dot{y} + F_y + \sqrt{2 \gamma K_b T} ( f_{y1} (t) + f_{y2} (t)),
\end{split}
\label{Eq_hor_1}
\end{equation}
where
\begin{equation}
    F_{x,y} = - 2U (L_{eq}) \frac{\partial S}{\partial d}\frac{\partial d}{\partial x,y},
\label{Eq_hor_2}
\end{equation}
describes the $x$ and $y$ components of the total restoring force with $d$ being the in-plane distance between the centers of the circular flakes of radius $R$, 
$S$ is the overlap area:
\begin{equation}
    S = \begin{cases}
    2 R^2 \operatorname{acos}( \frac{d}{2R} ) - d \sqrt{R^2 - (d/2)^2}, & |d| < 2R\\
    0, & |d| > 2R
    \end{cases} 
\label{Eq_hor_3}
\end{equation}
$f_{\nu,i}(t)$ are independent white noise terms describing the impact of thermal fluctuations of the reservoir on the flakes:
\begin{equation}
    \langle f_{\nu,i}(t) f_{\mu,i}(t - \tau) \rangle = \delta(\tau) \delta_{ij} \delta_{\mu \nu},
    \label{Eq_hor_4}
\end{equation}
and $\gamma_\parallel$ is the in-plane drag coefficient.


Figure \ref{fig9}(a) shows the resulting potential per unit area as a function of the in-plane center-to-center displacement $d$. The potential grows almost linearly near $d=0$ until it reaches the zero level when the overlap vanishes at $d > 2R$.

To calculate the in-plane Stokes' drag we employ the simple model of Couette flow \cite{landau2013fluid}:
\begin{equation}
    \gamma_\parallel = \frac{\eta \pi R^2}{L}
\label{Eq_hor_5}
\end{equation}
where $L = L_{eq}$ is the vertical distance between parallel flakes.
Figure \ref{fig9}(b) shows the corresponding diffusion coefficient $D$ as a function of the flake-to-flake distance $L$:
\begin{equation}
    D = \frac{K_B T}{\gamma_\parallel}.
\label{Eq_Diff}
\end{equation}
Despite the simplicity of the model, it reproduces very well numerically simulated diffusion coefficient for a similar kind of system obtained from rigorous hydrodynamic simulations \cite{schmidt2023tunable}.

Figure \ref{fig9}(c) presents a few simulated trajectories of the in-plane dynamics of the system for a pair of $R = 1~\mu$m circular flakes  performed for the CTAB concentration $C = 1$~mM at $T=300$~K. 
Trajectories feature Brownian dynamics with a characteristic displacement of a few tens of nanometers.
To study how this displacement if affected by the size of the flake, we run repeated simulations of Eq. \eqref{Eq_hor_1} for circular flakes of various radius and for each series of simulations estimate time-average center-to-center displacement $\langle d \rangle$. As Fig. \ref{fig9}(d) shows, the time-average center-to-center displacement reduces with increasing flake radius, which is due to the rapidly increasing in-plane Stokes' drag coefficient $\gamma_\parallel$, Eq. \eqref{Eq_hor_5}.

From the data presented so far one may also conclude that stable configuration of the system corresponding to the minimum of the total potential should also be stable against small rotation of the flakes around horizontal axes. 
The system is then no longer translationally invariant, and rigorous calculation of the Casimir potential would involve Green's tensor techniques \cite{Reid2009, Emig2007}.
Within the overlap approximation, however, for a small rotation angle we may roughly estimate the torque acting on both films by summing up the forces acting on the elementary segments of the films.
Indeed, consider the situation where one of the two identical films separated by the distance $L_{eq}$ experiences a fluctuation that causes it to rotate by a small angle around a horizontal axis.
Then the segments of the films that as the result of the initial fluctuation move away from each other will experience net attraction, while the segments that move towards each other will experience net repulsion. The resulting torque will tend to bring the system in the initial state with parallel flakes.

\section{Conclusion and Discussion}
To conclude, we have studied theoretically equilibrium configurations and their dynamical behavior in the system of two parallel metallic films coupled by the Casimir and electrostatic interactions.
Combination of the rigorous Lifhitz formalism for the Casimir potential and numerical molecular dynamics simulations for the electrostatic interaction leads to a simple analytical model of the system, 
allowing to describe the stationary equilibria of the system, as well as its stochastic behavior near those equilibria.
With use of this analytical model we have found the crucial role of the salt concentration on the equilibrium states of the system, as well as the effect of temperature.
Using the same model, we have analyzed the resonant mechanical frequencies of the self-assembled cavity, as well as its vertical and in-plane stochastic dynamics under the influence of thermal fluctuations of the environment.
The results demonstrate reasonable agreement with previously reported experimental data. 
The analytical model developed in this paper could be used for modelling of opto-mechanical properties of similar Casimir-force based self-assembled nanostructures.


\section*{Acknowledgments}
Authors acknowledge fruitful discussion with Timur Shegai and Oleg Kotov.
D.G.B. acknowledges the financial support from the Ministry of Science and Higher Education of the Russian Federation (No. 0714-2020-0002), Russian Science Foundation (23-72-10005), and the BASIS Foundation (grant 22-1-3-2-1).

\appendix

\section{Molecular dynamics simulations}

We performed dynamics simulations simulations of CTAB adsorption using GROMACS package \cite{lindahl2001gromacs}.To describe inter-atomic interactions we applied the OPLS-AA force field \cite{jorgensen1996development} with partial charges parameterized by LigParGen \cite{dodda2017ligpargen}. Interactions between CTAB and Au surface were described using parameters from GolP force-field, which shows good results for organics with gold \cite{iori2009golp}.   
The TIP4P \cite{jorgensen1983comparison} rigid non-polarizable model was used to parameterize the water molecules. A cutoff for short-range and non-bonded interactions was 1.2 nm. For long-range Coulomb interactions, we used the smooth Particle-Mesh Ewald scheme \cite{essmann1995smooth}. Visualization was produced using VMD \cite{humphrey1996vmd}.

\bibliography{Casimir}

\end{document}